\documentclass[aps,prb,floatfix,twocolumn,showpacs,reprint]{revtex4-1}
\usepackage{graphicx}
\usepackage{mathrsfs}
\usepackage{bm}
\usepackage{amsmath}
\usepackage{dcolumn}
\usepackage{epstopdf}
\usepackage{dsfont}
\usepackage{amssymb}
\usepackage{tabularx}
\usepackage{array}
\usepackage{color}
\usepackage[colorlinks=true, letterpaper=true, pdfstartview=FitV, linkcolor=blue, citecolor=blue, urlcolor=blue]{hyperref}

\begin{document}
\title{Electric control of topological phase transitions in Dirac semimetal thin films}

\author{Hui Pan}
\affiliation{Department of Physics, Beihang University, Beijing 100191, China}

\author{Meimei Wu}
\affiliation{Department of Physics, Beihang University, Beijing 100191, China}

\author{Ying Liu}
\affiliation{Research Laboratory for Quantum Materials and EPD Pillar, Singapore University of Technology and Design, Singapore 487372, Singapore}

\author{Shengyuan A. Yang}
\affiliation{Research Laboratory for Quantum Materials and EPD Pillar, Singapore University of Technology and Design, Singapore 487372, Singapore}

\begin{abstract}
  We investigate the effect of a vertical electric field on a Dirac semimetal thin film. We show that through the interplay between the quantum confinement effect and the field-induced coupling between sub-bands, the sub-band gap can be tuned and inverted, during which the system undergoes a topological phase transition between a trivial band insulator and a quantum spin Hall insulator. Consequently, one can electrically switch the topological edge channels on and off, making the system a promising platform for constructing a topological field effect transistor.
\end{abstract}


\maketitle

\section{Introduction}
The study of topological insulators (TIs) have been one of the most active research areas in the past ten years,\cite{Hasan2010,QiXL2011}
which revolutionized our understanding of the electronic band structure. It is now understood that there could be nontrivial topologies encoded in the electronic wave-functions, characterized by various topological invariants according to the symmetry class of the system, and physically manifested by the appearance of topological boundary states. For example, two-dimensional (2D) TIs, also known as the quantum spin Hall (QSH) insulators, are characterized by a $\mathbb{Z}_2$ topological invariant and have spin helical edge states on sample boundaries,\cite{Konig2007} for which back-scattering is suppressed in the presence of time reversal symmetry.\cite{Hasan2010,QiXL2011,Ando2013} Hence they hold great promise for applications such as low-dissipation electronics, spintronics, and quantum computations. For these TIs, the nontrivial topology as well as the boundary states are protected by the finite energy gap, i.e., they are robust against perturbations as long as the insulating gap does not close.

It was later realized that the topological classification could be pushed beyond insulators to states without a gap.\cite{Murakami2007,Wan2011,Burkov2011} In particular, a novel state called Dirac semimetal (DSM) has been proposed and successfully demonstrated in recent experiments for two crystalline materials Na$_3$Bi and Cd$_3$As$_2$.\cite{Young2012,WangZJ2012,WangZJ2013,LiuZK2014,ZhangY2014,Liu2014,Borisenko2014,Neupane2014,Jeon2014,Liang2015,Yi2014,Xu2015}
In these materials, the Fermi energy sits at two three-dimensional (3D) Dirac points---where the bands touch with a fourfold degeneracy---and the dispersion is linear along all three directions in reciprocal space. Each Dirac point can be viewed as consisting of two Weyl points of opposite chiralities and is protected by the crystalline symmetry.\cite{Young2012,Manes2012,WangZJ2012,WangZJ2013,Naga2014} Such unusual electronic structure endows the system with many intriguing properties like the surface Fermi arcs and the quantum magnetoresistance.\cite{Yi2014,Xu2015,Abri1998,Lund2014} Perhaps more importantly, DSMs are expected to be an ideal parent compound for realizing other novel topological states such as Weyl semimetals, TIs, and topological superconductors.\cite{LiuZK2014} Particularly in this regard, DSMs offer a simple alternative to achieve the 2D TI phase through the quantum confinement effect.\cite{WangZJ2013,Xiao2015} It has been shown that with increasing thickness, DSM thin films exhibit oscillations in the 2D $\mathbb{Z}_2$ invariant whenever a quantum well state crosses the Dirac point.\cite{WangZJ2013,Xiao2015} Hence a QSH phase can be realized by a proper control of the film thickness. Since the QSH phase has so far been detected in only a few systems, given its fundamental and technological importance, the new approach to realize it using DSMs would be of great interest. Furthermore, the unique properties of DSMs may offer new methods to manipulate the properties of the QSH phase.

Motivated by these recent breakthroughs and by the great interest in utilizing DSMs for topological devices, in this work, we investigate the possibility of electric control of the topological phase transitions in a DSM thin film. We show that by using a vertical electric field, DSM thin films can be switched between a topological QSH phase and a trivial insulator phase. Hence one can electrically manipulate the topological edge states, and the charge and spin conduction through a finite sample can be readily switched on and off.  This leads to a simple design of a DSM-based topological field effect transistor with advantages of fast-speed, low power consumption, and low dissipation, owing to the robust topological edge channels combined with full electric control.

Our paper is organized as follows. In Sec.~II, we discuss the modeling of a DSM thin film and we present an analytic derivation of its low energy levels and its topological phase under the influence of a vertical electric field.
In Sec.~III, we show the results of numerical calculation, which explicitly demonstrate the topological phase transition and the associated change in topological edge states and confirm the physical picture obtained from our analytical analysis.
Finally, we give a discussion of some aspects of the effect and its possible device applications, and summarize our results in Sec.~IV.

\section{Model and Analytic Analysis}
Our analysis is based on a generic low-energy effective model describing the DSMs A$_3$Bi (A=Na, K, Rb) and Cd$_3$As$_2$ as derived in previous works.\cite{WangZJ2012,WangZJ2013} In these materials, the states around Fermi energy can be expanded using a minimal four-orbital basis of $|S_{1/2},1/2\rangle$, $|P_{3/2},3/2\rangle$, $|S_{1/2},-1/2\rangle$, and $|P_{3/2},-3/2\rangle$. Around $\Gamma$-point in the Brillouin zone, the effective Hamiltonian expanded up to quadratic order in the wave-vector $k$ is given by
\begin{equation}\label{Heff}
\mathcal{H}(\bm k)=\varepsilon_0(\bm k)+\left[
                                \begin{array}{cccc}
                                  M(\bm k) & Ak_+ & 0 & 0 \\
                                  Ak_- & -M(\bm k) & 0 & 0 \\
                                  0 & 0 & M(\bm k) & -Ak_- \\
                                  0 & 0 & -Ak_+ & -M(\bm k) \\
                                \end{array}
                              \right],
\end{equation}
where $\varepsilon_0(\bm k)=C_0+C_1k_z^2+C_2(k_x^2+k_y^2)$, $k_\pm=k_x\pm ik_y$, and $M(\bm k)=-M_0+M_1k_z^2+M_2(k_x^2+k_y^2)$ with $M_0, M_1, M_2>0$ to reproduce the band inversion feature at $\Gamma$-point. The material-specific parameters $A$, $C_i$, and $M_i$ are determined by fitting the first-principles result or the experimental measurement. It has been shown that this model nicely captures the essential low-energy physics as compared with experiment.\cite{LiuZK2014,Liu2014,Jeon2014}

For bulk DSMs, model (\ref{Heff}) gives the energy dispersion $\mathcal{E}(\bm k)=\varepsilon_0(\bm k)\pm\sqrt{M(\bm k)^2+Ak_\|^2}$, where $\bm k_\|=(k_x, k_y)$ is the 2D wave-vector in the $k_x$-$k_y$ plane. The spectrum has two Dirac points located along the $k_z$-axis at $(0,0,\pm k_\text{D})$ with $k_\text{D}=\sqrt{M_0/M_1}$. Each Dirac point is four-fold degenerate and can be viewed as consisting of two Weyl nodes with opposite chiralities (as represented by the two $2\times 2$ diagonal sub-blocks in Hamiltonian (\ref{Heff})). The dispersion around each Dirac point is linear in all three directions, as can be seen by expanding $\mathcal{E}(\bm k)$ at $(0,0,\tau k_\text{D})$ ($\tau=\pm$ labels the two Dirac points): $\mathcal{E}(\bm k)\simeq \sqrt{A^2k_x^2+A^2k_y^2+4M_1^2k_\text{D}^2(k_z-\tau k_\text{D})^2}$. One notes that the low-energy spectrum is anisotropic as manifested in both the distribution of Dirac points as well as the different Fermi velocities along $k_z$ versus that in the $k_x$-$k_y$ plane (Fermi velocity along $k_z$ is typically much slower), which leads to quite different behaviors when a DSM is confined along different directions.\cite{Xiao2015}

DSMs such as Na$_3$Bi and Cd$_3$As$_2$ have layered structures along crystal $c$-axis. Hence their thin film structures with confinement along $z$-direction can be more readily fabricated. Consider a DSM thin film with thickness $L$ confined in the region $z \in[-L/2, L/2]$. For small $L$, the electron motion along $z$ will be quantized into discrete quantum well levels due to quantum confinement effect. This generally turns the system from a semimetal to a semiconductor. Using quantum well approximation, each quantum well level has a quantized effective wave-vector $k_{z}$ such that $\langle k_{z}\rangle_n=0$ and $\langle k_{z}^2\rangle_n\simeq (n\pi/L)^2$ with $n(=1,2,\cdots)$ counting the sub-bands and the angular bracket meaning the average over quantum well states.

One observes that for each sub-band $n$, Hamiltonian (\ref{Heff}) has a similar form as the low-energy model describing the 2D QSH systems in HgTe/CdTe quantum wells,\cite{Bern2006} with the term
\begin{equation}M(\bm k)\rightarrow M(n,\bm k_\|)=\mathcal{M}_n+M_2(k_x^2+k_y^2),
\end{equation}
where $\mathcal{M}_n\equiv -M_0+M_1(n\pi/L)^2$ is a sub-band dependent mass which determines the gap of the sub-band at $\Gamma$-point of the 2D Brillouin zone. It's known that band inversion occurs (around $\bm k_\|=0$) when $\text{sgn}(\mathcal{M}_n/M_2)=-1$,\cite{Bern2006,Shenbook} i.e. when $\mathcal{M}_n$ and $M_2$ have opposite signs, which signals a nontrivial $\mathbb{Z}_2=1$ character of the sub-band $n$. Given that $M_0,M_1,M_2>0$, this happens when $\tilde{k}_n\equiv \langle k_{z}^2\rangle_n^{1/2}(=n\pi/L)<k_\text{D}$ is satisfied. Therefore, for a thin-enough film such that $\tilde{k}_1=\pi/L>k_\text{D}$, all the sub-bands are topologically trivial with positive mass terms $\mathcal{M}_n$. With increasing film thickness, the system becomes nontrivial once the first ($n=1$) sub-band has its mass $\mathcal{M}_1$ inverted when $\tilde{k}_1<k_\text{D}$. The inverted sub-band contributes a $\mathbb{Z}_2=1$ and in the inverted band gap, there appears a pair of spin-helical edge states protected by time reversal symmetry on each edge of the quasi-2D system. Further increasing $L$ would invert $\mathcal{M}_2$ of the second sub-band, leading to two pairs of edge states. However, for $\mathbb{Z}_2$ group: $1+1=0$, hence this state is topologically trivial. Physically, it is because backscattering can occur between edge states from different time reversal pairs. Following this logic, the topological properties as well as the bulk band gap show oscillatory behavior as a function of the film thickness.\cite{WangZJ2013,Xiao2015}

Since the sub-band dependent mass $\mathcal{M}_n$ plays the key role in determining the topological properties of the system, we shall focus on the change of $\mathcal{M}_n$ by a vertical electric field, aiming to achieve an electric control of the topological phase of DSM thin films. To proceed, one notes that the lower diagonal block of Hamiltonian (\ref{Heff}) is formally the time reversal counterpart of the upper block, which share the same energy spectrum and the $E$ field does not mix the two.\cite{degeneracy} Hence to study the change of $\mathcal{M}_n$, it is enough to consider only the upper block
denoted by $h(\bm k)$. Modeling with the hard-wall boundary condition for the confinement potential, we have for sub-band $n$,
\begin{equation}
h_n(\bm k_\|)=\varepsilon_0(n,\bm k_\|)I+Ak_x\sigma_x-Ak_y\sigma_y+M(n,\bm k_\|)\sigma_z,
\end{equation}
where $\sigma$'s are the Pauli matrices, $I$ is the $2\times 2$ identity matrix, and $\varepsilon_0(n,\bm k_\|)=C_0+C_1(n\pi/L)^2+C_2 k_\|^2$. The energy eigenstates are given by
\begin{equation}
\langle\bm r|\Psi_{n\alpha}(\bm k_\|)\rangle=\frac{1}{\sqrt{S}}e^{ik_x x+ik_y y}\psi_n(z)\chi_{n\alpha},
\end{equation}
with eigen-energies
\begin{equation}
\mathcal{E}_{n\alpha}(\bm k_\|)=\varepsilon_{0}(n,\bm k_\|)+\alpha\sqrt{A^2k_\|^2+M(n,\bm k_\|)^2},
\end{equation}
where $S$ is the area of the thin film, $\alpha=\pm$, $\chi_{n\alpha}$ are the two eigen-spinors along the quantization direction $(Ak_x, -Ak_y, M(n,\bm k_\|))$, and
\begin{equation}
\psi_n(z)=\sqrt{\frac{2}{L}}\sin\left[\frac{n\pi}{L}(z+\frac{L}{2})\right]
\end{equation}
is the quantum well state for the $n$-th sub-band. The vertical electric field is modeled by adding a diagonal potential energy term $V(z)=eEz$ where $(-e)$ is the electron charge and $E$ is the effective field strength which may be considered as including the static screening effects.

For a qualitative analysis, we assume small field and treat $V$ perturbatively. Because $V(z)$ is odd in $z$, it is easy to see that the first order perturbation in energy vanishes. The leading order perturbation comes at the second order, with
\begin{equation}
\delta\mathcal{E}_{n\alpha}(\bm k_\|)\approx\sum_{m\beta\neq n\alpha}\frac{|\langle \Psi_{m\beta}(\bm k_\|)|{V}|\Psi_{n\alpha}(\bm k_\|)\rangle|^2}{\mathcal{E}_{n\alpha}(\bm k_\|)-\mathcal{E}_{m\beta}(\bm k_\|)},
\end{equation}
where the summation is over all other states different from $\Psi_{n\alpha}$, and in reality, it has a physical cutoff for which the low-energy description is no longer valid. One notes that in order to analyze the renormalized $\mathcal{M}_n$, it is sufficient to focus on the change at $\Gamma$-point of the 2D Brillouin zone by setting $\bm k_\|=0$.

We are most interested in the case in which the mass (gap) of the first sub-band can be inverted by the electric field, because then the two sides of the topological phase transition have the most salient contrast: absence or presence of the topological edge channels, hence leading to the best on-off ratio when considering a topological transistor based on it. For such case, we consider a thickness $L$ such that $\mathcal{M}_1=-M_0+M_1(\pi/L)^2>0$, i.e. an initially trivial system with $\mathbb{Z}_2=0$ in the absence of $E$ field. At $ k_\|=0$, we have $\chi_{n+}=|\uparrow\rangle$ and $\chi_{n-}=|\downarrow\rangle$ for all $n$, where $|\uparrow\rangle$ and $|\downarrow\rangle$ are the two eigenstates of $\sigma_z$. For the $n=1$ sub-band, we have at $ k_\|=0$,
\begin{equation}\begin{split}\label{delta1+}
\delta\mathcal{E}_{1+}&\approx -e^2E^2\sum_{m>1}\frac{|\langle \psi_m(z)|{z}|\psi_1(z)\rangle|^2}{\mathcal{E}_{m+}-\mathcal{E}_{1+}}\\
&=-e^2E^2\frac{64L^4}{\pi^6}\frac{1}{(M_1+C_1)}\eta.
\end{split}
\end{equation}
where
\begin{equation}\label{eta}
\eta=\sum_{m\in\text{even}}\frac{m^2}{(m^2-1)^5}.
\end{equation}
In the first equality of (\ref{delta1+}) we used the fact that the state $|\Psi_{1+}(0)\rangle$ does not mix with the $|\Psi_{m-}(0)\rangle$ states from the valence bands by the $E$ field because their pesudo-spin parts $\chi$ are orthogonal. Also note that one has $M_1>|C_1|$ in order for the model (\ref{Heff}) to describe a semimetal phase, hence $\mathcal{E}_{m+}>\mathcal{E}_{1+}$ for $m>1$, hence the perturbation due to the coupling between $|\Psi_{1+}(0)\rangle$ and $|\Psi_{m+}(0)\rangle$ (with $m>1$) generally pushes down the energy level of $|\Psi_{1+}(0)\rangle$, making $\delta\mathcal{E}_{1+}<0$. In the expression (\ref{eta}) of the constant factor $\eta$ (with a rapidly converging value $\simeq 0.0165$), the summation only includes the even integer numbers, because $V(z)$ only couples states with opposite parities in $z$.

Similarly, the energy shift for state $|\Psi_{1-}(0)\rangle$ can be calculated,
\begin{equation}
\delta\mathcal{E}_{1-}\approx e^2E^2\frac{64L^4}{\pi^6}\frac{1}{(M_1-C_1)}\eta,
\end{equation}
which is positive, showing that the coupling induced by the $E$ field pushes up the energy of $|\Psi_{1-}(0)\rangle$. Therefore, in the Hilbert sub-space of the first sub-band, the $E$ field renormalizes the value of the mass:
\begin{equation}
\mathcal{M}_1\rightarrow \mathcal{M}_1+\delta\mathcal{M}_1,
\end{equation}
with the correction
\begin{equation}
\delta\mathcal{M}_1(E)=\frac{\delta\mathcal{E}_{1+}-\delta\mathcal{E}_{1-}}{2}\approx -e^2E^2 \frac{64L^4}{\pi^6}\frac{2M_1}{M_1^2-C_1^2}\eta.
\end{equation}
Using this estimation, one observes that the gap of the first sub-band would decrease with increasing $E$ field, and closes at
\begin{equation}\label{Ec}
E_c\approx \frac{\pi^3}{8e^2L^2}\sqrt{\frac{M_1^2-C_1^2}{2\eta M_1}\mathcal{M}_1},
\end{equation}
which signals a topological phase transition point and beyond which the gap reopens with the system turned into a QSH insulator phase characterized by $\mathbb{Z}_2=1$.

For large $E$ field with $eEL$ being comparable or even larger than $\mathcal{M}_1$, the perturbative calculation is no longer expected to be accurate. Nevertheless, the general physical pictures from the above discussion still applies: the level repulsion due to higher sub-bands would generally decrease and invert the gap of the first sub-band, generating a topological phase transition. We shall explicitly demonstrate this in the next section through numerical calculations.

The above analysis can also be applied to higher sub-bands when a thicker film with the $n$-th ($n>1$) sub-band most close to transition is considered. We will discuss this later in Sec.~IV.

\section{Numerical Results}

For numerical investigation, we discretize the model (\ref{Heff}) on a 3D lattice with lattice constants $a_x=a_y=0.5448$ nm ($1.264$ nm), and with $a_z=0.4828$ nm ($2.543$ nm) being set to the interlayer separation pertinent to Na$_3$Bi (Cd$_3$As$_2$). The standard substitutions
\begin{equation}
k_i\rightarrow \frac{1}{a_i}\sin(k_i a_i),\qquad k_i^2\rightarrow \frac{2}{a_i^2}[1-\cos(k_i a_i)]
\end{equation}
are adopted ($i=x,y,z$) for lattice discretization around $\Gamma$-point. Since we require the initial state at $E=0$ is of a trivial insulator phase, we need the number of layers $\ell<\lfloor \pi M_1/(M_0 a_z)\rfloor +1$, where $\lfloor\cdots\rfloor$ is the floor function. And in order for the band gap to be inverted by a relatively small $E$ field, one may wish to have the initial gap size $2\mathcal{M}_1$ not too large.

\begin{figure}
  \includegraphics[width=8.5cm]{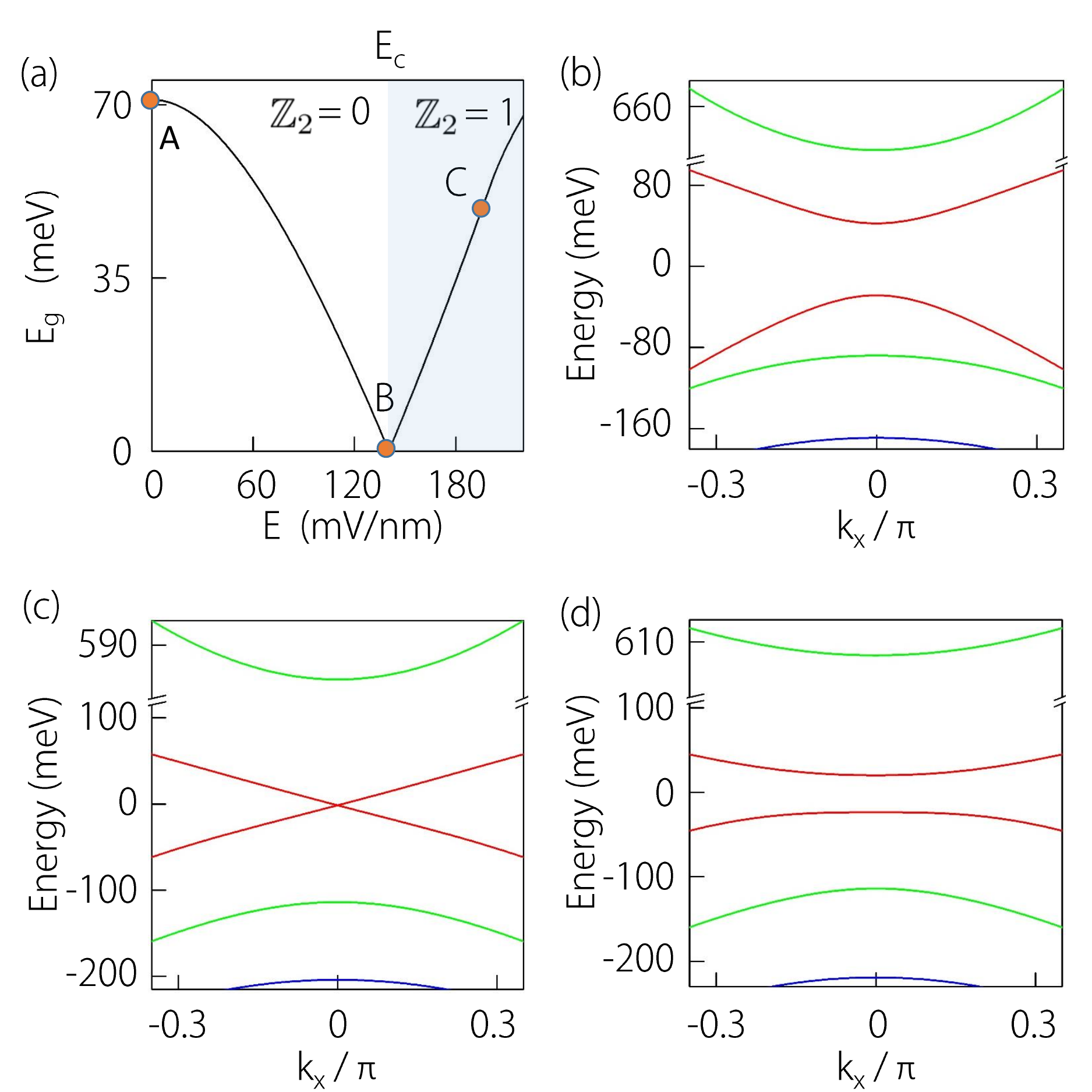}
  \caption{(color online) Field induced topological phase transition in Na$_3$Bi thin film. (a) Variation of energy gap as a function of the vertical field $E$ showing the gap closing and reopening process, marking a topological phase transition between $\mathbb{Z}_2=0$ and $\mathbb{Z}_2=1$ phases. (b-d) Energy spectra corresponding to A, B, and C as marked in (a) plotted versus $k_x$ (with $k_y=0$). The first, second, and third sub-bands are marked using red, green, and blue colors respectively.  The parameters for used in the calculation are $\ell=5$, $C_0=-63.82$ meV, $C_1=87.536$ $\mathrm{meV}\cdot \mathrm{nm}^2$, $C_2=-84.008$ $\mathrm{meV}\cdot \mathrm{nm}^2$, $M_0=86.86$ meV, $M_1=106.424$ $\mathrm{meV}\cdot \mathrm{nm}^2$, $M_2=103.610$ $\mathrm{meV}\cdot \mathrm{nm}^2$, and
  $A=245.98$ $\mathrm{meV}\cdot \mathrm{nm}$. For better comparison, a rigid energy shift is applied to make the gap center at zero energy. \label{FIG:NaBiEz}}
\end{figure}

Let's consider Na$_3$Bi first. The model parameters we use are listed in the caption of Fig.~\ref{FIG:NaBiEz}, which have been extracted from the first-principles calculations and compared well with experiment. For Na$_3$Bi thin films, the critical thickness for which the first sub-band undergoes band inversion is around $\ell_c\simeq 6 $.\cite{Xiao2015} Hence we take a film with $\ell=5$ layers ($L\simeq2.414$ nm) for demonstration. Fig.~\ref{FIG:NaBiEz}(a) shows the variation of the band gap $E_\text{g}$ as a function of the $E$ field. The result is symmetric between positive and negative values of $E$, so only the positive part is shown here. Initially, at $E=0$, the system has a confinement gap about $71$ meV (marked by point A). With increasing $E$ field, the gap decreases and closes at a critical value $E_c\simeq140$ mV/nm (marked by point B), and then reopens and increases with $E$. This is consistent with our previous discussion in Sec.~II. The value of $E_c$ is also not far from our estimation in Eq.(\ref{Ec}) which is about 208 mV/nm. We also plot in Fig.~\ref{FIG:NaBiEz}(b-d) the energy spectrum of the system corresponding to the three representative states marked by A, B, and C in Fig.~\ref{FIG:NaBiEz}(a). It shows that on both sides of the gap closing point, the system is insulating with the gap belong to the first sub-band. At the critical value $E_c$, the band gap closes at $\Gamma$-point, marking the topological phase boundary which separates the topologically trivial and nontrivial phases.

\begin{figure}
  \includegraphics[width=8.6cm]{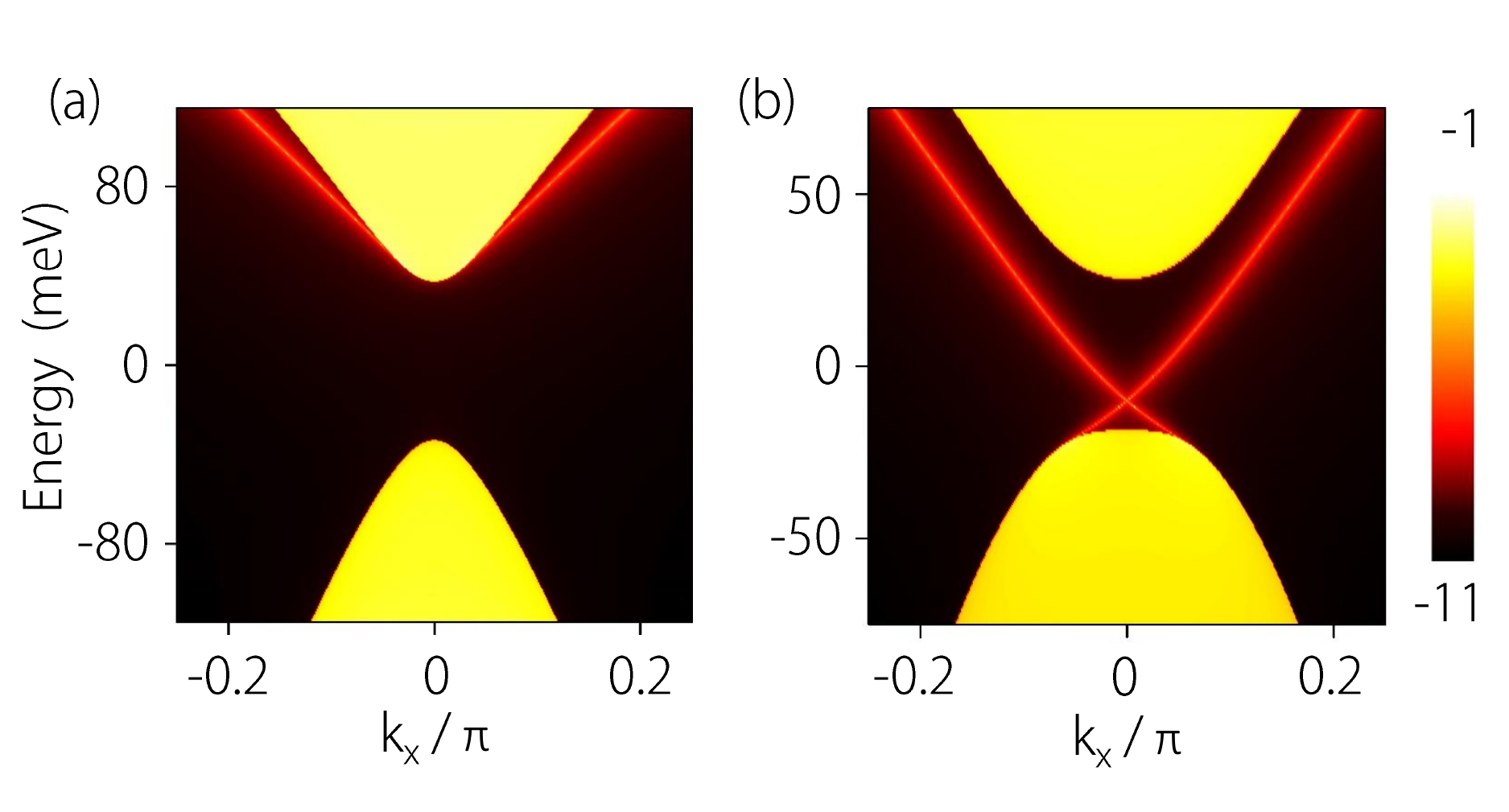}
  \caption{\label{FIG:NaBiLODS} (color online) Calculated LDOS of a side surface for Na$_3$Bi thin film. (a) is for point A (trivial insulator) and (b) is for point C (QSH insulator) as marked in Fig.~\ref{FIG:NaBiEz}(a). The calculation is for a slab which is semi-infinite along $y$-direction and the parameters are the same as for Fig.~\ref{FIG:NaBiEz}.}
\end{figure}

To further demonstrate the topological nature of the transition and to visualize the edge states, we compute the surface local density of states (LDOS) for the side surface. Due to the isotropy in the $k_x$-$k_y$ plane of the low-energy model (\ref{Heff}), without loss of generality, we choose the surface perpendicular to $y$-direction of the quasi-2D system. The surface LDOS $\rho(k_x)$ can be calculated for each $k_x$ from the surface Green's function $\rho(k_x)=-\text{Tr}[\text{Im}G_{00}(k_x)]/\pi$, where $G_{00}$ is the retarded Green's function for the surface layer (labled by index $0$) of the lattice.\cite{Chu2011} $G_{00}$ can be evaluated by the transfer matrix through a standard numerical iterative method.\cite{Sanc1984} The obtained surface LDOS for states before and after the phase transition (for state A and C) are plotted in Fig.~\ref{FIG:NaBiLODS}.
One observes that for both cases, the confinement-induced bulk gap can be clearly identified. Before the topological phase transition ($E<E_c$), there is no states inside the gap. In contrast, after transition ($E>E_c$), there appear two bright lines crossing the gap, corresponding to the spin helical edge states for the $\mathbb{Z}_2$ nontrivial QSH phase. As long as time reversal symmetry is preserved, these gapless modes are protected and carriers in these channels cannot be backscattered.\cite{Hasan2010,QiXL2011} Therefore transport through these channels is in principle dissipationless.

\begin{figure}
  \includegraphics[width=6.8cm]{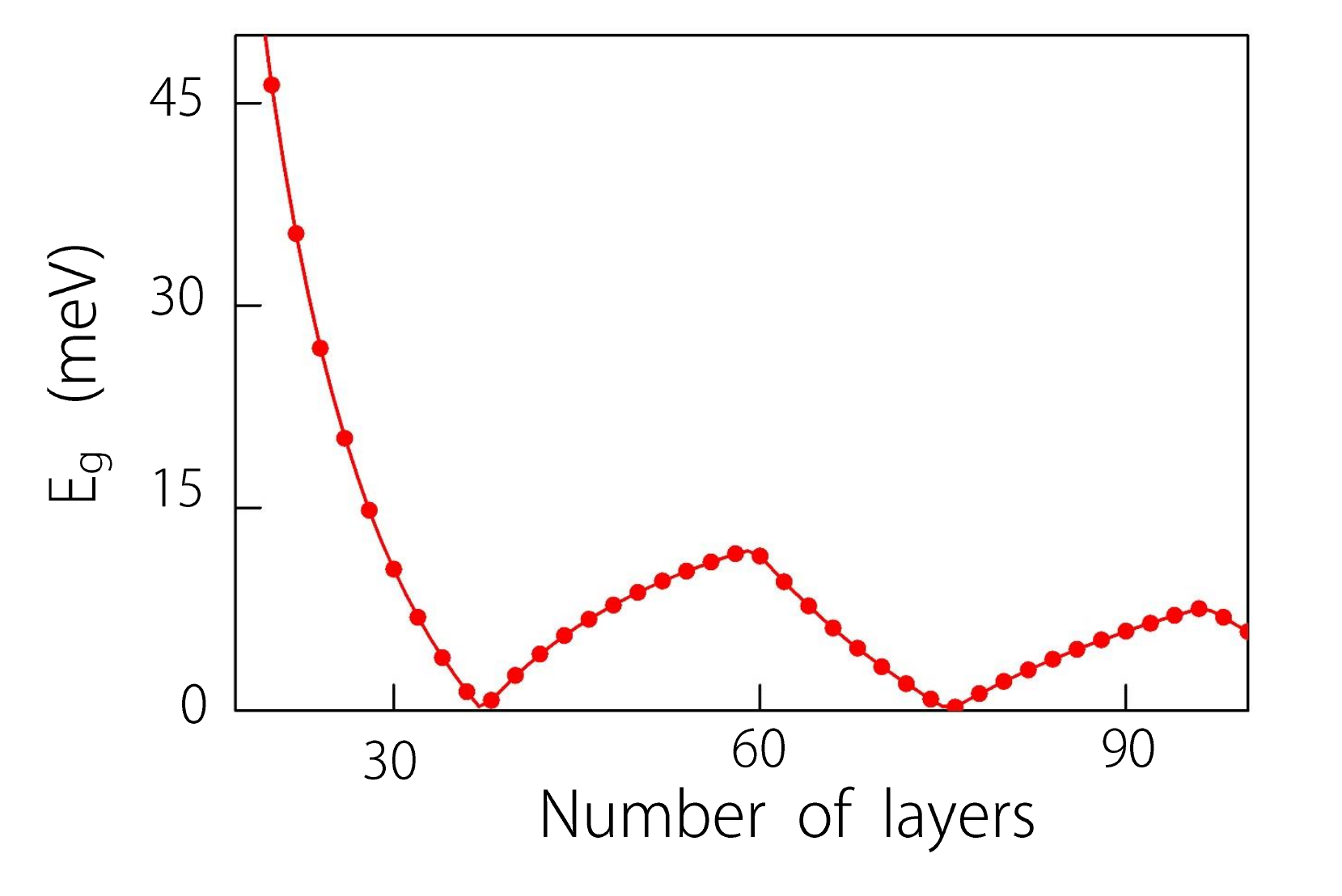}
  \caption{\label{FIG:CdAsNcell} (color online)  Confinement induced gap size versus thickness $\ell$ for Cd$_3$As$_2$ thin films. The parameters  used in the calculation are $C_0=-219$ meV, $C_1=-300$ $\mathrm{meV}\cdot \mathrm{nm}^2$, $C_2=-160$ $\mathrm{meV}\cdot \mathrm{nm}^2$, $M_0=10$ meV, $M_1=9600$ $\mathrm{meV}\cdot \mathrm{nm}^2$, $M_2=180$ $\mathrm{meV}\cdot \mathrm{nm}^2$, and $A=275$ $\mathrm{meV}\cdot \mathrm{nm}$.}
\end{figure}

\begin{figure}
  \includegraphics[width=8.5cm]{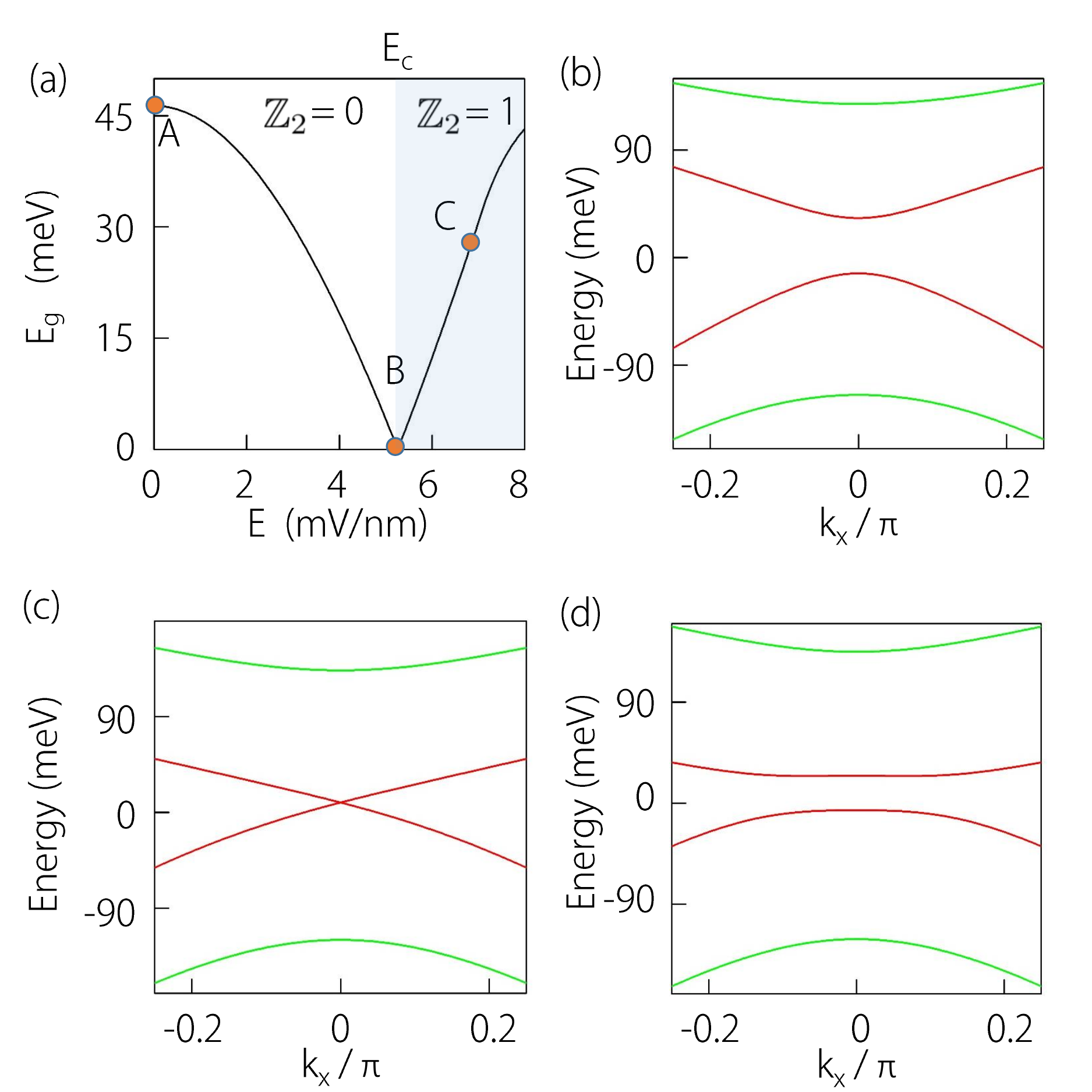}
  \caption{\label{FIG:CdAsEz} (color online) Field induced topological phase transition in Cd$_3$As$_2$ thin film. (a) Variation of energy gap as a function of the vertical field $E$. (b-d) Energy spectra corresponding to A, B, and C as marked in (a) plotted versus $k_x$ (with $k_y=0$). The first and the second sub-bands are marked using red and green colors respectively. The parameters for used in the calculation are the same as for Fig.~\ref{FIG:CdAsNcell} and $\ell=20$ is taken. }
\end{figure}

\begin{figure}
  \includegraphics[width=8.6cm]{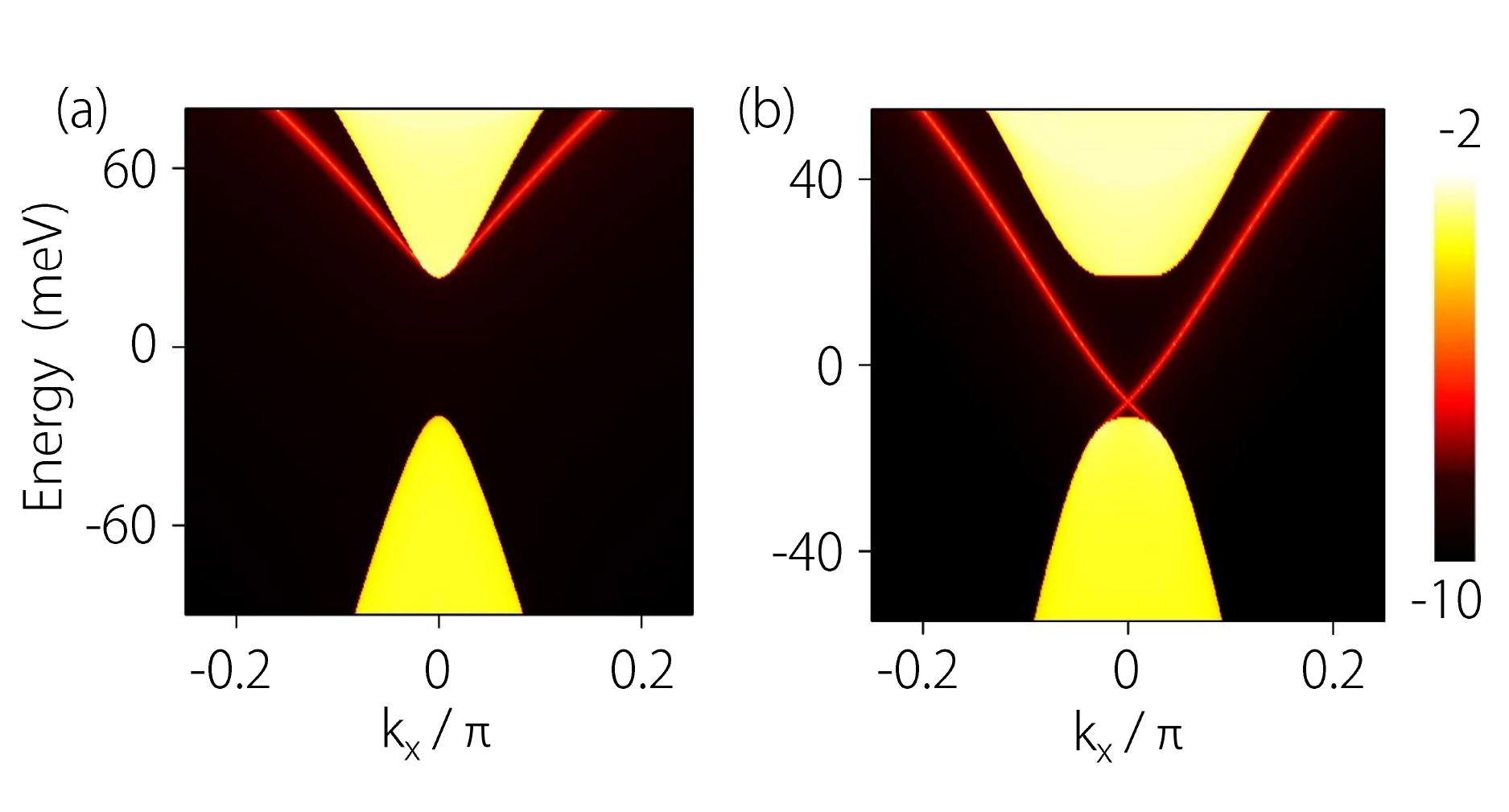}
  \caption{\label{FIG:CdAsLDOS} (color online) Calculated LDOS of a side surface for Cd$_3$As$_2$ thin film. (a) is for point A (trivial insulator) and (b) is for point C (QSH insulator) as marked in Fig.~\ref{FIG:CdAsEz}(a). The calculation is for a slab which is semi-infinite along $y$-direction and the parameters are the same as for Fig.~\ref{FIG:CdAsEz}.}
\end{figure}

Similar analysis applies to Cd$_3$As$_2$ as well. In Fig.~\ref{FIG:CdAsNcell}, we plot the variation of its confinement-induced gap versus the film thickness, which clearly shows the oscillation behavior of the gap.\cite{Xiao2015} One observes that the critical thickness $\ell_c$ is at about $37$ layers. Here we choose a film thickness of $\ell=20$ layers ($L=50.86$ nm) for demonstration. The variations of the gap with respect to the $E$ field as well as representative energy spectra are shown in Fig.~\ref{FIG:CdAsEz}. Again the gap closing and reopening process similar to Fig.~\ref{FIG:NaBiEz}(a) is observed. The critical value of $E_c\simeq5.26$ mV/nm also agrees well with the estimation $\approx 4.17$ mV/nm from Eq.(\ref{Ec}). The energy spectra also coincide with our previous analysis. Fig.~\ref{FIG:CdAsLDOS} shows the side surface LDOS plots for states A and C (marked in Fig.~\ref{FIG:CdAsEz}(a)), clearly showing the appearance of topological edge states across the transition. These results show qualitatively the same features as those for Na$_3$Bi.

Our numerical results discussed above thus confirm our analytical analysis in Sec.~II. A vertical electric field can be used to control the topological phase transitions and the topological edge states in a DSM thin film.

\section{Discussion and Summary}

This work theoretically demonstrates the possibility to electrically control the $\mathbb{Z}_2=0/1$ topological phase transitions in a DSM thin film. Since the bulk topology is tied to the existence of topological edge channels and hence to the charge/spin conductance, this indicates that one can achieve a full electric control of the on/off charge/spin conductance of such a system, making it a suitable candidate for a topological field effect transistor. The electric field can be generated by the standard top and bottom gates setup. Compared with the traditional MOSFET which works by injection and depletion of charge carriers in the channel region and has a response timescale depending on factors such as the charge concentration and the carrier mobility, the operating mechanism for a topological transistor is expected to have a high on/off speed with electronic response timescale and a better power efficiency.\cite{Qian2014} In addition, multiple conducting channels in a transistor can be achieved by designing a multilayer structure with alternating DSM layers and insulating layers, similar to the structure as in Ref.\onlinecite{Qian2014}.

For device design, we have seen that a proper film thickness can be chosen such that the starting confinement gap is small hence can be easily inverted by a small applied field. However, there is a tradeoff because if the gap is too small, then the thermally populated carriers in the bulk could strongly contribute to the transport. Therefore, a balance needs to be achieved for the device to have an optimal performance with relatively low power consumption.

In our analysis, we have focused on the phase transition in the first sub-band. Similar analysis can also be extended to higher sub-bands if a thicker film with its $n$-th ($n>1$) sub-band close to phase transition is considered. For example, consider a film thickness such that its second sub-band is just before the gap-closing. In this case, we would have $\mathcal{M}_2>0$ and $\mathcal{M}_1<0$, and the system is in a $\mathbb{Z}_2=1$ phase. Perturbation to second order in the field strength gives the energy correction of
\begin{multline}\label{2nd}
\delta\mathcal{E}_{2+}\approx -e^2E^2\frac{256 L^4}{\pi^6}\left(\frac{1}{M_1+C_1}\eta'\right.\\
\left.-\frac{1}{81[3C_1+5M_1-2M_0(L/\pi)^2]}\right),
\end{multline}
for the $|\Psi_{2+}(0)\rangle$ state of the second sub-band, with $\eta'=\sum_{m\in\text{odd},m\geq 3}{m^2}/{(m^2-4)^5}$. A similar expression can be obtained for $\delta\mathcal{E}_{2-}$ as well. The first term in the parenthesis of (\ref{2nd}) is from the coupling with the $m>2$ sub-bands, while the second term is from the coupling with the first sub-band. The sign of this energy shift (and hence the correction of $\mathcal{M}_2$) would depend on the competition between the two terms and is not necessarily negative. Nevertheless, for such higher sub-band case, even if a field-induced topological phase transition can be realized, the topologically trivial phase would in fact still possess edge states. Although these (even number of pairs of) channels are not topologically robust, their presence would make the trivial state not completely `off' hence is detrimental to the performance of a transistor.

In summary, we have demonstrated that full electric control of topological phase transitions in a DSM thin films can be achieved through the interplay between the quantum confinement effect and the coupling between sub-bands induced by a vertical electric field. As a result, a topological field effect transistor can be constructed in which carriers are conducted through the topological edge channels. Given that several DSM materials have been experimentally demonstrated and that the progress in material fabrication technology such as molecular beam epitaxy has allowed film growth with atomic precision, it is quite promising for the physical effect and the DSM-based topological transistor proposed here to be realized in the near future.

{\color{blue}\emph{Acknowledgement.}} The authors would like to thank D.L. Deng for helpful discussions. This work was supported by NSFC under Grant No. 11174022, NCET program of MOE, and SUTD-SRG-EPD2013062.


\begin{thebibliography}{999}
\bibitem{degeneracy} This can also be argued by observing that the low-energy Hamiltonian possesses the unitary symmetry $\mathcal{U}=I\otimes \sigma_z$ and the anti-unitary symmetry $\mathcal{U}'=-i \sigma_y\otimes I K$, which may be regarded as emergent symmetries for the low-energy physics. Here $\sigma$'s are the Pauli matrices, $I$ is the $2\times 2$ identity matrix, and $K$ is the complex conjugation operator.
\end{thebibliography}
\end{document}